\documentclass{iaus}
\usepackage{graphicx}
\newcommand{\mnras}{MNRAS}
\newcommand{\aap}{A\&A}

\newcommand{\apj}{ApJ}

\newcommand{\pasp}{PASP}

\title[N-body simulations of star clusters] 
{N-body simulations of star clusters}

\author[Anders, Lamers \& Baumgardt]   
{Peter Anders$^1$, Henny J.G.L.M. Lamers$^1$ \and Holger Baumgardt$^2$}

\affiliation{$^1$Sterrenkundig Instituut, Universiteit Utrecht, P.O. Box 80000, 3508 TA Utrecht, The
Netherlands; email: anders/lamers@astro.uu.nl  \\[\affilskip]
$^2$ Argelander-Institut f\"ur Astronomie, Auf dem H\"ugel 71, 53121 Bonn, Germany \break 
email: holger@astro.uni-bonn.de}

\pubyear{2007}
\volume{246}  
\pagerange{119--126}
\date{?? and in revised form ??}
\jname{Dynamical Evolution of Dense Stellar Systems}
\editors{E. Vesperini, M. Giersz, \& A. Sills, eds.}
\begin{document}

\maketitle

\begin{abstract}
Two aspects of our recent N-body studies of star clusters are presented:\\
1) What impact does mass segregation and selective mass loss have on integrated photometry?\\
2) How well compare results from N-body simulations using NBODY4 and STARLAB/KIRA?
\end{abstract}

\firstsection 
\section{Selective mass loss and integrated photometry}

``Mass segregation'' describes the effect that high-mass stars are preferentially found
in the center of clusters, while the outskirts are preferentially occupied by lower-mass
stars.

(Primordial) mass segregation is studied observationally in a number
of clusters. Some examples are: the Orion Nebula  Cluster
(\cite{1998ApJ...492..540H}), 6 LMC clusters
(\cite{2002MNRAS.331..245D}).

Dynamical mass segregation is also found from N-body simulations (e.g.
\cite{1985ApJ...292..339I,1997MNRAS.286..709G}), caused by two-body
encounters and energy equipartition. A number of authors also point at
the preferential mass loss of low-mass stars when the cluster evolves
in a tidal field (e.g.
\cite{1975ApJ...201..773S,1997MNRAS.286..709G}), as mass segregation
populates the cluster outskirts preferentially with low-mass stars
where they are most easily stripped from the cluster potential.  The
resulting changes in the overall stellar mass function inside the
cluster are quantified by \cite{2003MNRAS.340..227B}.

In \cite{2006A&A...452..131L} we use the results from \cite{2003MNRAS.340..227B} to
incorporate the changing mass function slope in a simplified manner into the GALEV code
(see \cite{2003A&A...401.1063A} and references therein) to calculate its impact on the
integrated photometry of star clusters.

Our main findings are:
\begin{itemize}

\item at 0 -- 40 per cent of the cluster's lifetime: the cluster colours are comparable to
standard models (i.e. without preferential loss of low-mass stars)

\item at 40 -- 80 per cent of the cluster's lifetime: a cluster appears too blue/young
(compared to standard models) due to the loss of lower main-sequence stars 

\item at 80 -- 100 per cent of the cluster's lifetime: a cluster appears too red/old (compared
to standard models) due to the loss of main sequence turn-off stars

\item when interpreting photometry of mass-segregated clusters, that have preferentially lost
low-mass stars, with standard photometric
models (without taking mass-segregation effects into account) the derived ages can be
wrong by 0.3 -- 0.5 dex

\end{itemize}

\section{Benchmark test for N-body codes: Comparing NBODY4 with STARLAB/KIRA}

Here we propose a benchmark test for comparison of N-body codes: 1024
equal-mass particles, Plummer sphere model, without primordial
binaries, external tidal field or stellar/binary evolution.

\begin{center}
\begin{figure}
	\vspace{-0.5cm}
	\hspace{1.2cm}
	\begin{tabular}{cc}
		\includegraphics[angle=270,width=0.4\columnwidth]{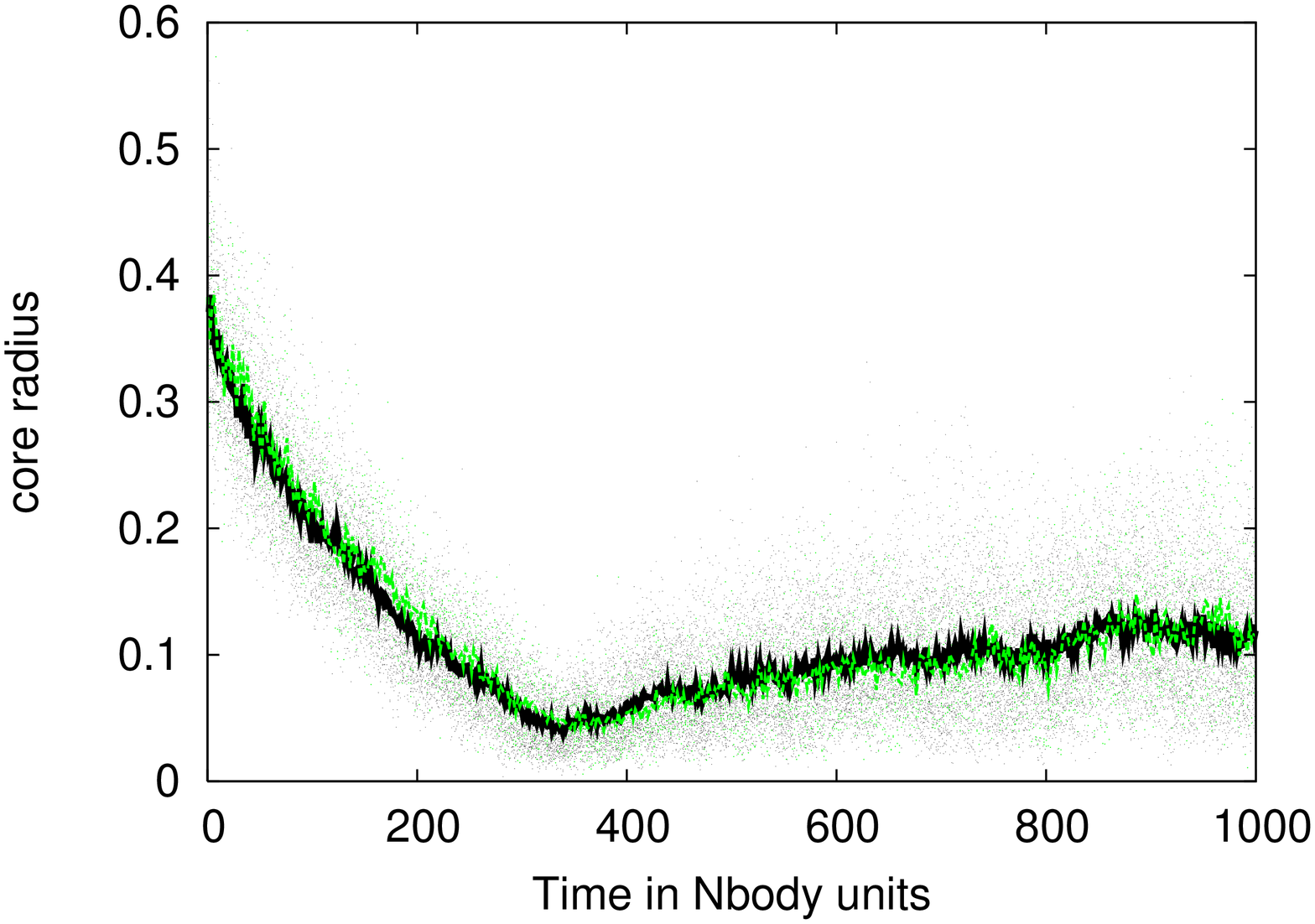} &
 		\includegraphics[angle=270,width=0.4\columnwidth]{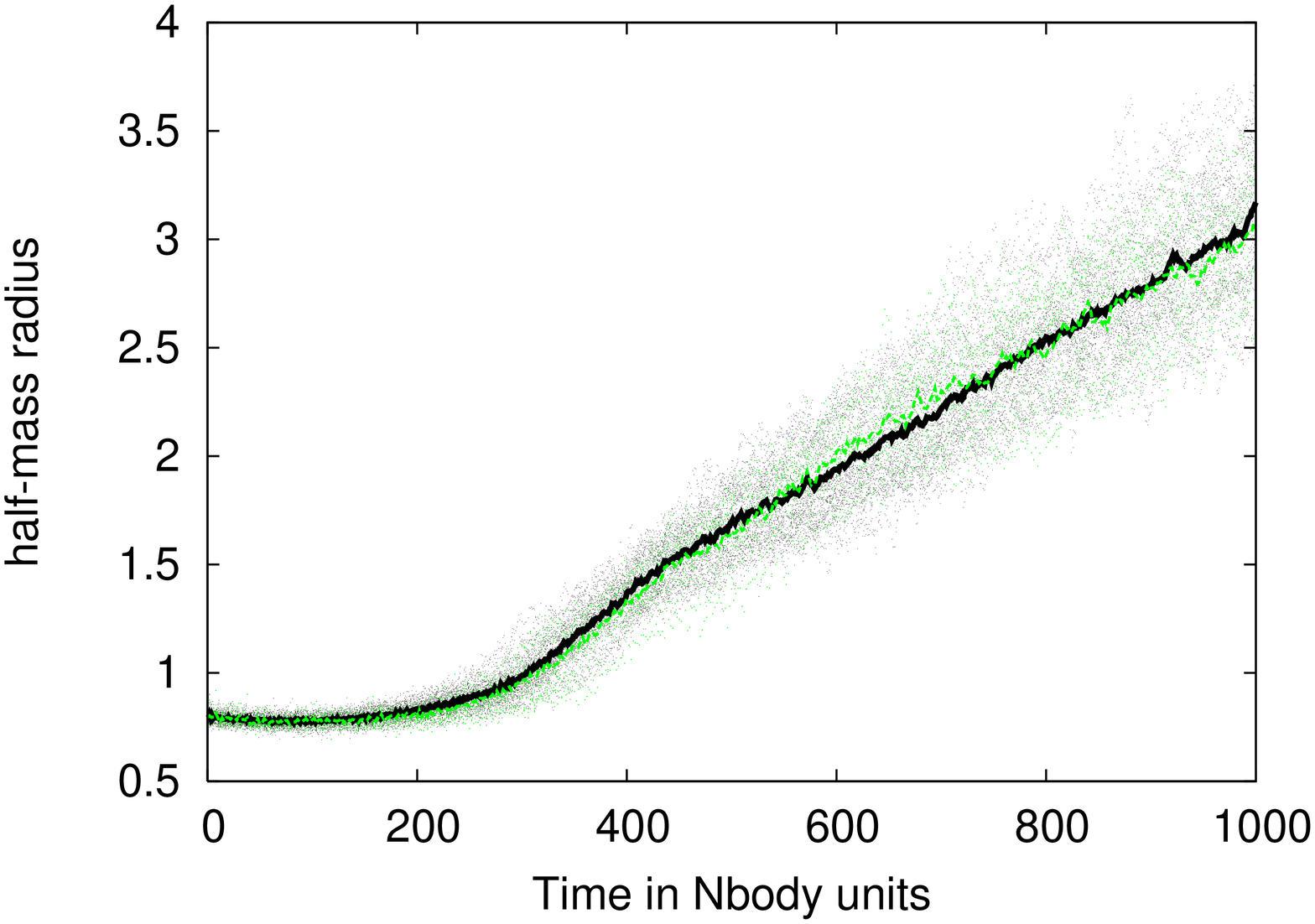} \\	
 		\includegraphics[angle=270,width=0.4\columnwidth]{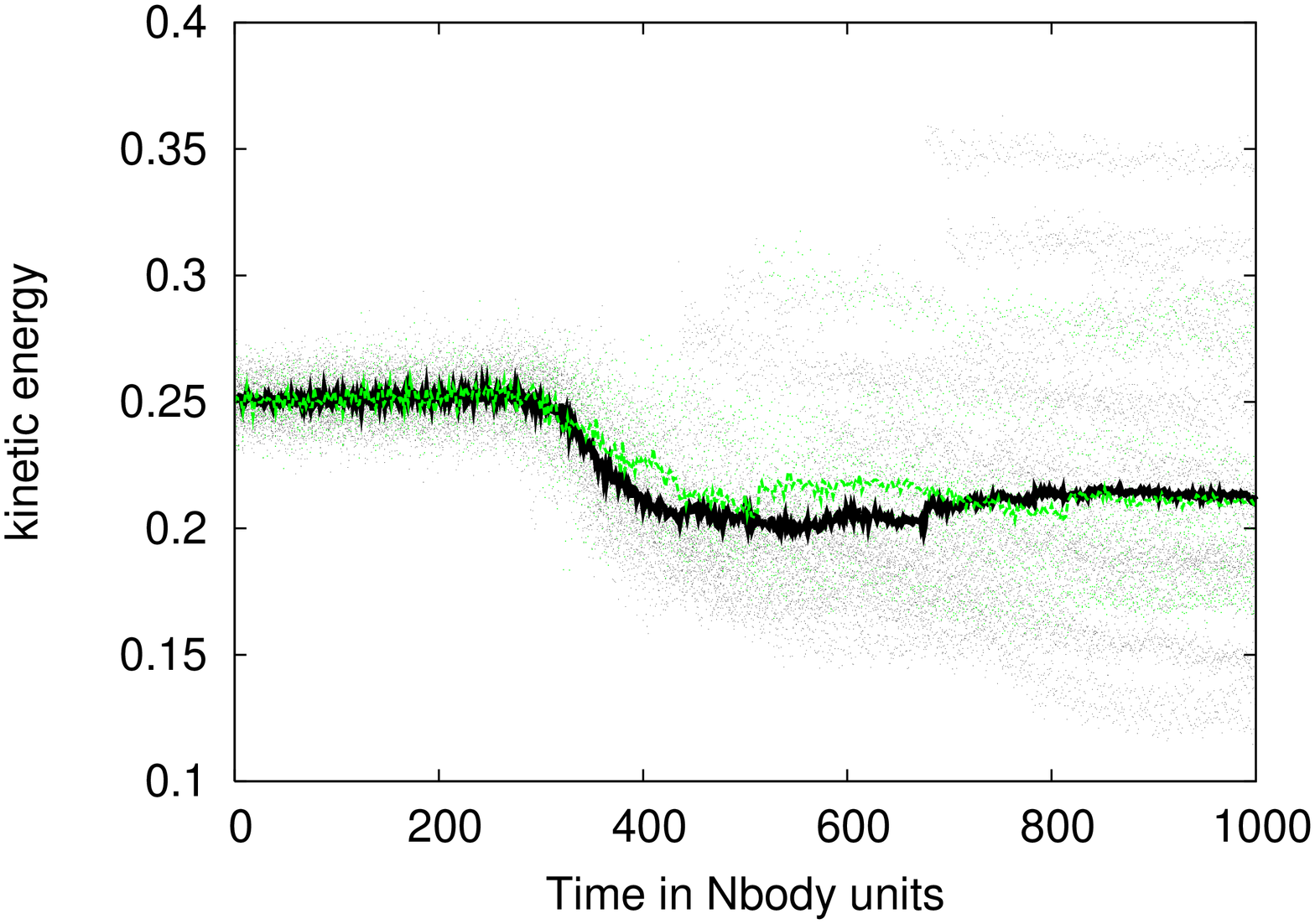} &
 		\includegraphics[angle=270,width=0.4\columnwidth]{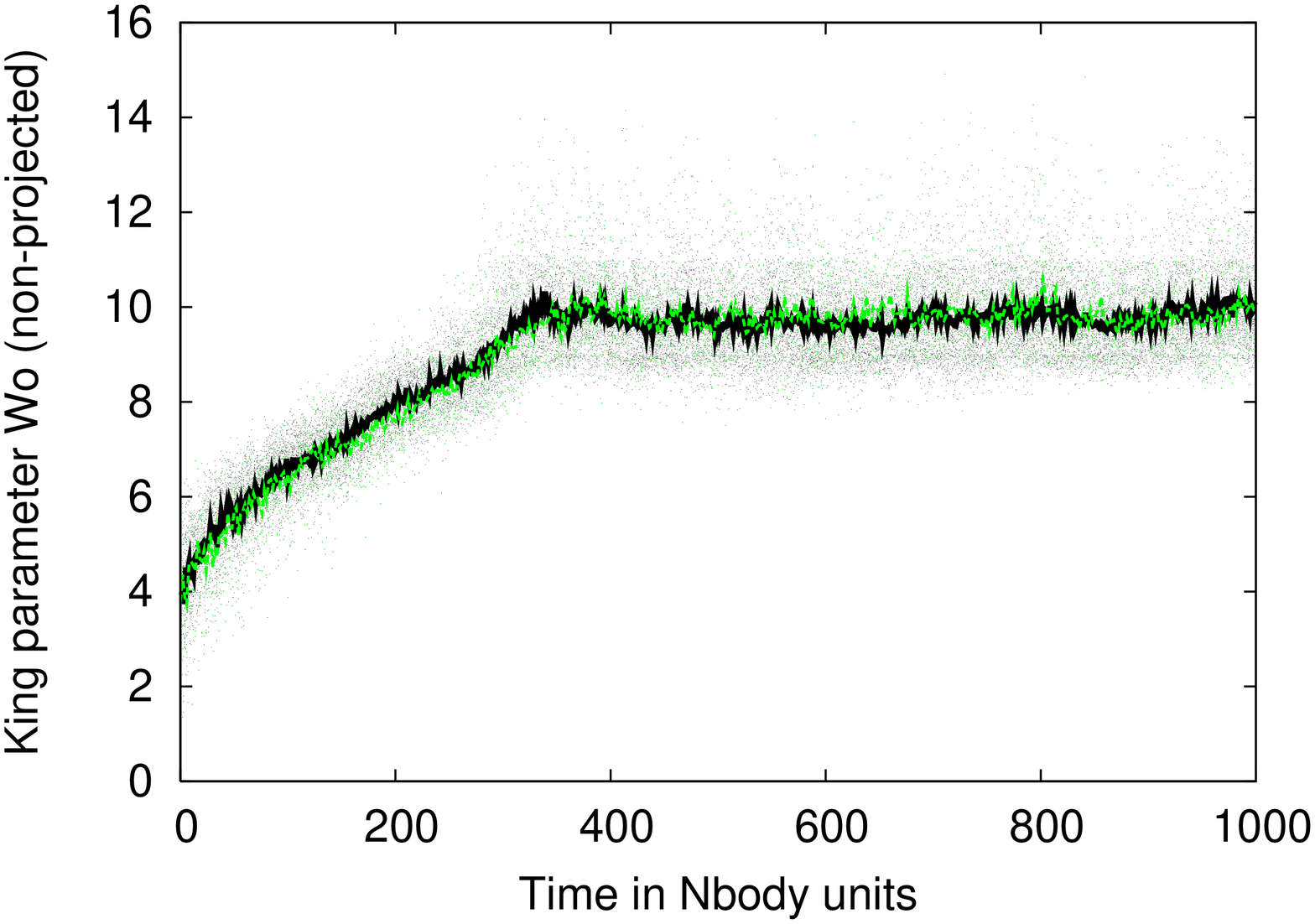} \\
	\end{tabular}
	\caption{Comparison of simulations using STARLAB (black) vs NBODY4 (green/grey). The
	dots represent the individual runs, the lines show the mean values.}
	\label{fig1}
\end{figure}
\end{center}

We performed simulations using the same input files (10 individual
runs) for NBODY4 (see e.g. \cite{aarseth99}) and STARLAB (see e.g.
\cite{simon01}), following the setup from the proposed benchmark test.
We find stochastic effects to be of importance for 10 runs, hence we
supplemented the STARLAB results with 20  additional runs (equivalent
NBODY4 runs are in preparation). Some results are shown in Fig.
\ref{fig1}. For most parameters studied the results are virtually
indistinguishable. Some questions remain for the kinetic energy and
the parameter  distributions of dynamically formed binaries. Any
differences there could be originating from the different treatment of
binaries, or due to stochastical effects. Further tests (e.g. with
primordial binaries) are in preparation.

\vspace{-0.5cm}

\begin{acknowledgments}
We thank the ISSI in Bern,
Switzerland, where parts of this project were carried out. 
We thank Evghenii Gaburov and Simon
Portegies Zwart for comments and suggestions.
\end{acknowledgments}


\vspace{-0.5cm}

\begin{thebibliography}

\bibitem[Aarseth 1999]{aarseth99} Aarseth, S.\ 1999, \pasp, 111, 1333

\bibitem[Anders \& Fritze-v.~Alvensleben 2003]{2003A&A...401.1063A} 
Anders, P., \& Fritze-v.~Alvensleben, U.\ 2003, \aap, 401, 1063 


\bibitem[Baumgardt \& Makino (2003)]{2003MNRAS.340..227B} Baumgardt,
H., \&  Makino, J.\ 2003, \mnras, 340, 227 


\bibitem[de Grijs et al.2002]{2002MNRAS.331..245D} de Grijs, R., Gilmore, 
G.~F., Johnson, R.~A., \& Mackey, A.~D.\ 2002, \mnras, 331, 245 


\bibitem[Giersz \& Heggie 1997]{1997MNRAS.286..709G} Giersz, M., \& 
Heggie, D.~C.\ 1997, \mnras, 286, 709 


\bibitem[Hillenbrand \& Hartmann 1998]{1998ApJ...492..540H} Hillenbrand, 
L.~A., \& Hartmann, L.~W.\ 1998, \apj, 492, 540 


\bibitem[Inagaki \& Saslaw 1985]{1985ApJ...292..339I} Inagaki, S., \& 
Saslaw, W.~C.\ 1985, \apj, 292, 339 

%
%

\bibitem[Lamers et al. (2006)]{2006A&A...452..131L} Lamers, H.~J.~G.~L.~M., 
Anders, P., \& de Grijs, R.\ 2006, \aap, 452, 131 

\bibitem[Portegies Zwart et al. 2001]{simon01} Portegies Zwart,
S.~F., McMillan, S.~L.~W., Hut, P., \& Makino, J.\ 2001, \mnras, 321,
199

\bibitem[Spitzer \& Shull 1975]{1975ApJ...201..773S} Spitzer, L., Jr., \& 
Shull, J.~M.\ 1975, \apj, 201, 773 

\end{thebibliography}
\end{document}